\documentclass[letter]{aa} 

\titlerunning{Stellar streams around dwarf galaxies}
\authorrunning{J. D. Sakowska et al}

\usepackage{pdflscape}
\usepackage{txfonts}
\usepackage{graphicx}

\usepackage{natbib}
\usepackage{gensymb}

\bibpunct{(}{)}{;}{a}{}{,}
\usepackage[colorlinks=true, linkcolor=blue, citecolor=blue, urlcolor=blue]{hyperref}

\begin{document}

   \title{Stellar streams around dwarf galaxies in the Local Universe}
   
   \subtitle{}

   \author{Joanna D. Sakowska,
          \inst{1}
          David~Mart\'inez-Delgado\inst{2,3}\fnmsep\thanks{ARAID fellow}, Sarah Pearson\inst{4}, Francisco J. Riquel-Castilla\inst{5}, Tjitske K. Starkenburg\inst{6,7}, Giuseppe Donatiello\inst{8}, Alis Deason\inst{9}, Denis Erkal\inst{10}, Ethan D. Taylor\inst{10}
          }
\institute{
$^{1}$ Instituto de Astrofísica de Andalucía (CSIC), Glorieta de la Astronom\'\i a,  E-18080 Granada, Spain\\
$^{2}$ Centro de Estudios de F\'isica del Cosmos de Arag\'on (CEFCA), Unidad Asociada al CSIC, Plaza San Juan 1, 44001 Teruel, Spain\\
$^{3}$ ARAID Foundation, Avda. de Ranillas, 1-D, E-50018 Zaragoza, Spain\\
$^{4}$ Niels Bohr International Academy $\&$ DARK, Niels Bohr Institute, University of Copenhagen, Blegdamsvej 17, 2100 Copenhagen, Denmark\\
$^{5}$ Facultad de Físicas, Universidad de Sevilla, Avda. Reina Mercedes s/n, Campus Reina Mercedes, E-41012 Seville, Spain\\
$^{6}$ CIERA and Department of Physics and Astronomy, Northwestern University, 1800 Sherman Ave, Evanston, IL 60201, USA \\
$^{7}$ NSF–Simons AI Institute for the Sky (SkAI), 172 E. Chestnut St., Chicago, IL 60611, USA\\
$^{8}$UAI -- Unione Astrofili Italiani /P.I. Sezione Nazionale di Ricerca Profondo Cielo, 72024 Oria, Italy \\
$^{9}$ Institute for Computational Cosmology, Department of Physics, Durham University, South Road, Durham DH1 3LE, UK\\
$^{10}$ Department of Physics, University of Surrey, Guildford GU2 7XH, UK \\
}

   \date{Received 28 November 2025; accepted 30 January 2026}

  \abstract
   {While mergers between massive galaxies and their dwarf satellites are well studied, the properties of dwarf -- dwarf satellite mergers are not well constrained. Stellar streams trace satellite disruption and, in the dwarf galaxy regime, are predicted to provide novel constraints on low-mass galaxy evolution and dark matter. However, the mass ratios required to form these streams make them challenging to detect.}
   {We present a preview of the Stellar Stream Legacy Survey (SSLS) in the dwarf galaxy regime. The SSLS aims to produce a statistically large, homogeneous sample of stellar streams for comparison with galaxy evolution theory.}
   {We visually inspect dwarf galaxies using the DESI Legacy Imaging Survey (DES and DECaLS footprints, $r$-band$\sim$29 mag arcsec$^{-2}$) within 4 -- 35 Mpc. We develop a classification metric to categorise accretion debris around dwarf galaxies, and measure the frequency of accretion features in the DES footprint only.}
   {We present the first release of accretion features around dwarf galaxies collected from the DES and DECaLS footprints, including 1 stream, 11 shells, and 8 asymmetric stellar halos, of which 17 constitute new identifications. In the DES footprint, we inspect 730 dwarfs and find that 5.1$\%$ (37/730) show accretion features. Although this frequency measurement is lower than the SSLS result for massive galaxies, we discuss the observational biases behind detecting streams in the dwarf galaxy regime.}
   {Our results highlight the difficulty of detecting streams around dwarfs, and identify the need for improved theoretical modelling of low-mass merger morphologies. Nevertheless, they place constraints on hierarchical mass assembly in this regime.}


   \maketitle
   
\nolinenumbers

\section{Introduction}

\begin{figure*}
\begin{center}
\includegraphics[width=1\textwidth]{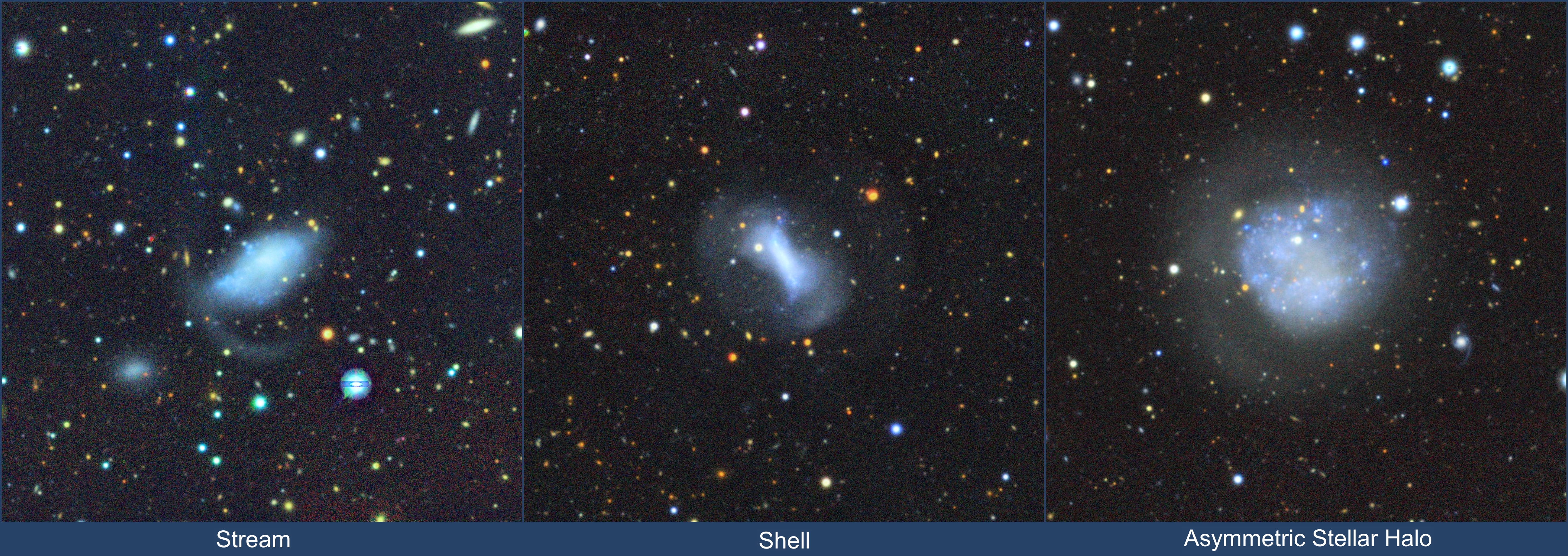}
\caption{Morphological classification of satellite accretion in dwarfs, resulting from our inspection of the DES and DECaLS footprints: stream (left, ESO 508-509), shell (middle, PGC 40604) and asymmetric stellar halo (right, PGC 46382).
}\label{fig:morphology}
\end{center}
\end{figure*}

In the Lambda-Cold Dark Matter framework ($\Lambda$CDM), galaxies grow hierarchically via the accretion of smaller systems (e.g., \citealt{White1978}). Dwarf galaxies ($M_{\star} < 10^{9.5} M_{\odot}$) should therefore have hosted their own satellite systems (\citealt{Wheeler2015, Dooley2017b})- even if they have already been tidally destroyed by their host (\citealt{diemand2008}) or are dark matter (DM) halos (e.g. \citealt{Helmi2012, Starkenburg2015, Sawala2015}). Testing these predictions is crucial for understanding structure formation, as small-scale structure represents one of the most significant observational challenges for $\Lambda$CDM (e.g., \citealt{boylan-kolchin2017}).

Stellar streams represent ideal tracers of past satellite accretion, preserving the archaeological record of merger events long after the original satellite has been completely disrupted. For massive galaxies ($M_{\star} \geq 10^{10} M_{\odot}$), tidal debris from infalling satellites (i.e. minor mergers with total mass ratios $\leq$1:3, \citealt{Mancillas2019}) survives for Gyrs (\citealt{Johnston2001, Cooper2013}) exhibiting diverse morphologies, including long stellar streams, shells, jet-like features, and giant debris clouds (e.g., \citealt{MD2023, Sola2025}). The observed properties (morphology, surface brightness [SB], colours) and frequency of streams can be cross-compared with predictions from both cosmological and zoom-in simulations of MW-like galaxies, and data-to-model tensions identified (\citealt{MC2025}). By expanding the census of known streams, we can obtain statistical constraints on their host galaxies' DM halo shapes from morphologies alone (\citealt{Nibauer2023}), as well as constraints on the halo masses and radial profiles (\citealt{Pearson2022, Walder2024, Nibauer2025}). Extending these techniques to streams around dwarf galaxies will be particularly powerful for constraining DM properties. Dwarfs are highly DM-dominated systems, and their merger frequencies and stellar halo properties are highly sensitive to assumed DM particle physics (\citealt{Deason2022}). As alternative DM simulations become more readily available, e.g. DREAMS (\citealt{Rose2025, Lin2024}), FIRE self-interacting DM (SIDM; \citealt{FIRESIDM}), EAGLE warm and SIDM (\citealt{EAGLESIDM}), AIDA-TNG (\citealt{Despali2025}), streams will play an important role in deciphering whether solutions to data-to-model tensions require modifications in the dark sector.

While we know of \textit{at least} 100+ extragalactic streams around MW–mass hosts, significantly less progress has been made in identifying streams around dwarf galaxies. Beyond our Local Group, only a handful of dwarfs with stellar streams are known (e.g., NGC 4449, DDO 68, NGC 300; \citealt{Martinez-Delgado2012, Annibali2016, Fielder2025}). Targetted campaigns, such as the Smallest Scale of Hierarchy survey (SSH; \citealt{Annibali2020}), are resolving accretion signatures in stellar halos up to the edge of the Local Volume (<11 Mpc) in 45 isolated dwarfs (\citealt{Annibali2022, Pascale2022, Pascale2024, Sacchi2024}). Yet the frequency of hierarchical mass assembly in the low-mass regime is not well understood. Previous works have catalogued merging dwarfs (z < 0.02; \citealt{Paudel2018}), or measured the frequency of tidal debris around isolated dwarfs (z < 0.12; \citealt{kado-fond2020}), but no frequency measurement of streams and shells (distinguished from tidal arms) has been presented. Such a result would enable cross-examination with future predictions made by cosmological simulations on detectable accretion features in the low-mass regime.

The ultimate goal of the Stellar Stream Legacy Survey (SSLS, \citealt{MD2023, MC2024, MC2025}) is to probe $\Lambda$CDM cosmological simulations against a statistically significant stream sample, in order to identify discrepancies between galaxy formation theory and the observed stream properties and frequency. This Letter constitutes the first step towards this goal in the low-mass regime. In Section \ref{sec:meth}, we present our data and visual inspection method. In Section \ref{sec:res}, we discuss our new discoveries, placing conclusions in Section \ref{sec:conc}.

\section{Data and method}
\label{sec:meth}
\begin{figure*}
\begin{center}
\includegraphics[width=0.95\textwidth]{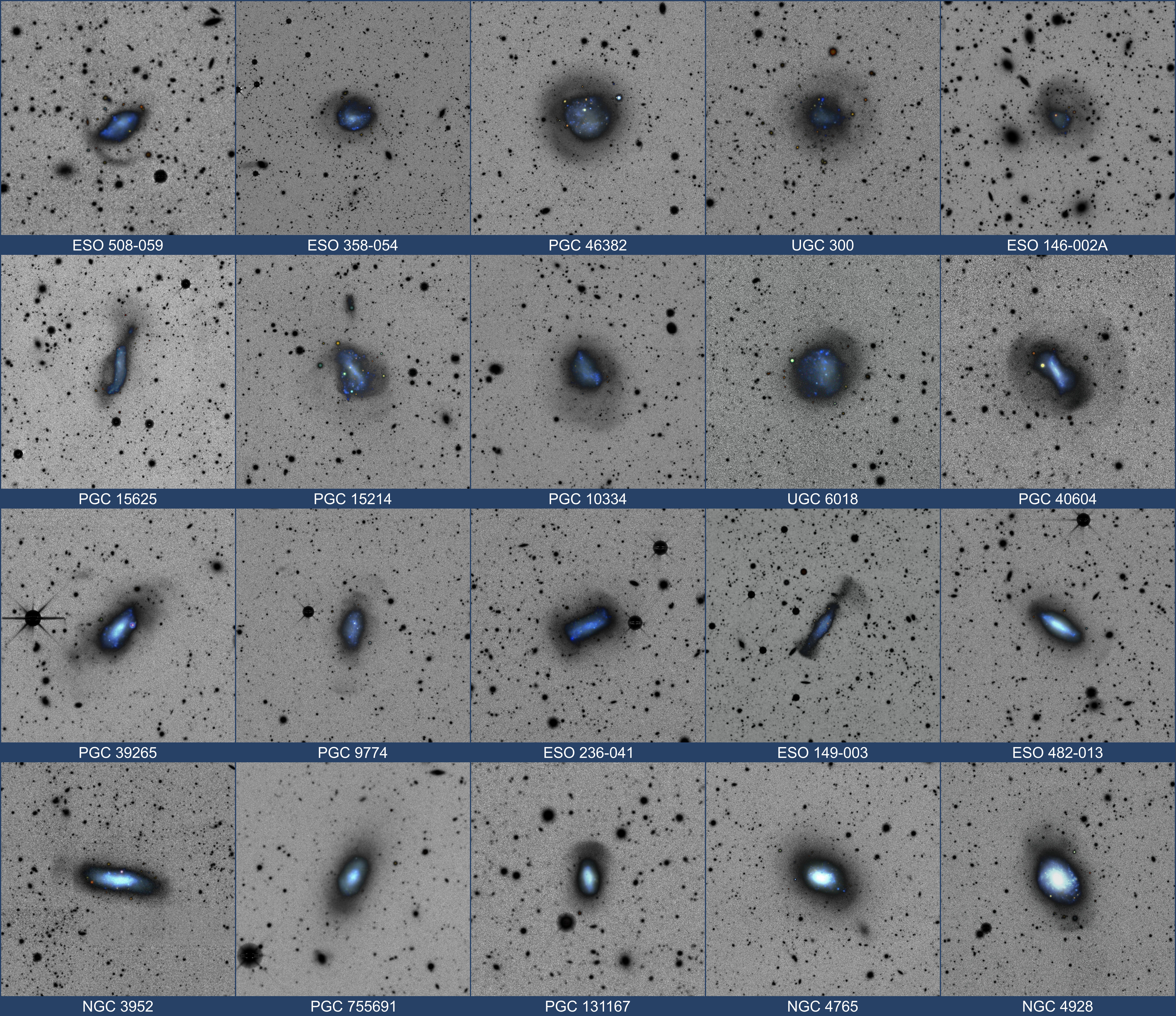}
\caption{Examples of accretion features around dwarf galaxies, classified with the metric in Figure \ref{fig:morphology} and ordered by morphology. We select 10 examples from DES, and 10 examples from DECaLS. We find a stream (upper left, ESO 508-509), asymmetric stellar halos (ESO 358-054 to UGC 6018, horizontally) and shells (PGC 40604 to NGC 4928, horizontally).}\label{fig:mosaic}
\end{center}
\end{figure*}

Our aims consist of (i), a proof-of-concept pilot, manual visual inspection of the SSLS footprint covered by the Dark Energy Camera (DES, DECaLS) to create a satellite accretion classification metric (objective 1), and (ii), obtaining the frequency of this event through a quantitative, manual visual inspection of the DES footprint only, guided by this metric (objective 2). 

We curate two (DES, DECaLS) dwarf galaxy samples by crossmatching the 50 Mpc Galaxy Catalog (50MGC; \citealt{Ohlson2024}) to the footprints. The 50MGC catalogue provides homogenized mass, distance and morphological type measurements for 15,424 nearby galaxies from three sources: HyperLeda (\citealp{Makarov2014}), the Local Volume Galaxy (LVG) catalogue (\citealp{Karachentse2013}) and the NASA-Sloan Atlas (NSA)\footnote{https://www.sdss.org/dr13/manga/manga-target-selection/nsa/}. Dwarf galaxies are selected with stellar mass $M_{\star} < 10^{9.5} M_{\odot}$, has a distance between $ 4 - 35$ Mpc and galactic latitude $|b| > 30\degree$, using the \verb|bestdist|, \verb|logmass| columns. For each dwarf, we calculate the projected separation $D_{project}$ and velocity separation $v_{sep}$ (using the \verb|v_cmb| column) from other galaxies in the 50MGC catalogue. Only dwarfs with $D_{project} > 400$ kpc and $|v_{sep}| > 250$ km s$^{-1}$ from a massive galaxy ($M_{\star} \geq 10^{10} M_{\odot}$) are selected. Dwarf galaxy pairs are flagged ($D_{project} < 50$ kpc, $|v_{sep}| < 250$ km s$^{-1}$) and discarded during the visual inspection (next step). This is not done automatically as we take care to not remove disrupted, smaller satellite progenitors that are part of an accretion feature. However, we do remove dwarfs with a smaller, intact satellite that has (at least) a distance measurement. We do not remove satellites with no catalogue entry, given it is unclear of their distance and association. Future SSLS releases will implement a tailored isolation criteria to fine-tune the results for comparison with predictions of observable dwarf accretion features from cosmological simulations, akin to \cite{MC2025} for MW-mass hosts, once they become available. The sample is manually inspected for accretion features using the DESI Legacy Imaging Survey (DESI LS\footnote{https://www.legacysurvey.org/viewer}) DR10 images. DESI LS includes other survey data that provide additional information on each galaxy (e.g., we used GALEX to check for recent star formation, sometimes indicative of tidal arms). We take care to not mistake spiral arms as accretion features, e.g., bright blue arms extending from the galactic disc, or dusty spiral arms coalescing after dwarf post-mergers.

Figure \ref{fig:morphology} shows our morphological classification of accretion features around dwarf galaxies across DES and DECaLS (objective 1): i) stellar streams, which are not part of the disc (ESO 508-059); ii) clearly discernable shells (PGC 40606) and iii) asymmetric stellar halos (PGC 46382). We now apply this metric to classify dwarf galaxies in DES to obtain the frequency measurement (objective 2). During the visual inspection of 934 entries we remove 204 contaminants. This includes dwarf galaxy pairs, dwarfs with intact satellites (see previous step), dwarfs infalling into a massive host (which where not removed automatically, given missing \verb|v_cmb| entries for the satellite), edge-on spiral galaxies, background redshifted galaxies, background post major mergers (massive galaxies), coalescing post mergers that were not clearly dwarfs, catalogue entries with no DES data. In summary, we examine 730 DES dwarf galaxies for accretion features, guided by the Figure \ref{fig:morphology} metric.

\section{Results and discussion}
\label{sec:res}

In Figure \ref{fig:mosaic}, we select 10 positive cases each from DES and DECaLS to present the most spectacular examples of accretion features discovered in our proof-of-concept study (objective 1). We order them by apparent morphology and show a stream (ESO 508-059), asymmetric stellar halos (ESO 358-054 horizontally to UGC 6018) and shells (IC 700 through NGC 4928). 17 of these 20 cases are new discoveries, with only PGC 40604, ESO 149-003, and NGC 4765 previously known. We discuss each case individually in Appendix \ref{appendix:A}. 

We find only one stream morphology in the entire survey (DECaLS footprint) and none in DES. Within the DES footprint, 5.1$\%$ (37) of dwarf galaxies show an accretion feature (shell/asymmetric stellar halo), including the 10 shown in Figure \ref{fig:mosaic}. We find 15 very low-mass ($M_{\star} \sim 10^{7} - 10^{8} M_{\odot}$) cases. In the same DES footprint, and using the same data and SB limits, \cite{MC2024} found $9.1\% \pm 1.1\%$ of MW-mass galaxies to have an accretion feature (stream, shell, and other morphologies). While this could indicate genuine differences for merger rates and debris morphologies across mass regimes, several observational factors complicate direct comparison. First, dwarf minor mergers ($<$1:10 total mass) may not contribute enough stars to appear in stellar halo SB profiles. The stellar mass -- halo mass relation at these scales is poorly known and depends on the assumed halo mass threshold for galaxy formation, which is not established. Major mergers ($>$1:3 total mass) should be more detectable as, not only do they bring in more stars, but they can also form extended distributions in the stellar halo (due to `pulled-out' host material) affecting SB profiles (\citealt{Deason2022}). Second, dwarf major mergers might preferentially form shells. While formation depends primarily on the progenitor orbit (radial versus circular/polar), there are secondary effects depending on mass. \cite{Amorisco2017} has shown that dynamical friction can act to radialize orbits when the accreted dwarf is relatively massive. Scaling this to dwarf mergers, one could argue that major mergers are more likely to have radial orbits and create shells. Taken together, shells from major mergers could be more structurally stable, and thus detectable for longer, than streams. Indeed, the stability of streams formed by dwarf mergers, and how long they last before they phase mix and create asymmetric stellar halos (like those in Figure \ref{fig:mosaic}), is not yet constrained. Finally, the detectability and classification bias of streams versus shells should be explored, given we can only make classifications based on the brightest pieces of debris. Dwarf tidal debris is already at the observable limit (e.g., \citealt{Deason2022, Tau2025}), and the viewing angle is expected to affect the apparent SB and even classification- for instance, `edge-on' streams will appear brighter (and potentially shell-like) in comparison to `face-on' streams. The effect of the viewing angle on the detectability of streams will be explored as future work (Sakowska et al., in prep). In summary, we highlight the difficulty in finding streams around dwarfs, and find tentative evidence for a shell detection bias.

We found it challenging to unambiguously separate accretion from morphological features without further theoretical information on the dwarf-dwarf merger sequence. During the inspection we identified a variety of complex dwarf morphologies, including 20 dwarf post-mergers with spiral arms still mixing (which resembled accretion tracks). We discuss these observational challenges in Appendix \ref{appendix:B}. For this reason, we consider 5.1$\%$ to be an upper limit to the observable number of accretion features in the DES sample at this SB-limit. Improving our statistic requires an extensive library of $N$-body hydrodynamical simulations spanning the dwarf-dwarf merger sequence. Such simulations have successfully reproduced faint substructures seen in nearby (< 11 Mpc) isolated dwarf galaxies (e.g., DDO 68, NGC 5238, UGC 8760, \citealt{Pascale2022, Pascale2024}) and 1:10 dwarf pairs (\citealt{Besla2016, Pearson2018}). Ideally, this library would also account for SB-limits and viewing angle effects on stream appearance and brightness.

\section{Conclusions}
\label{sec:conc}

This Letter comprises the first SSLS release of 20 dwarf galaxies with accretion features (1 stream, 11 shells, and 8 asymmetric stellar halos) identified by visual inspection of the DES/DECaLS footprints within 4 -- 35 Mpc. In the DES footprint only, using the same data and SB-limits as for our previous studies, we find that dwarfs show comparatively less ($5.1\%$) accretion features than MW-like galaxies ($9.1\% \pm 1.1\%$). An extensive $N$-body hydrodynamical model library of dwarf-dwarf satellite mergers is needed to better disentangle accretion from morphological features and constrain our result, which we regard as an upper limit. The observability of these extremely faint features must be investigated further before confirming a physical difference in the merger rates. We should assess whether we are observationally biased towards detecting major mergers, and if this is an important driver behind the dominance of shells in our survey.
Looking forward, our survey will provide dwarf galaxy stream candidates for follow-up studies- e.g. DM halo fitting, a technique expected to place novel constraints on DM candidates. The frequency measurement provided will test upcoming predictions made by cosmological simulations on observable accretion features in low-mass galaxies. Finally, our work previews what will be achievable with upcoming wide-field surveys, i.e. Euclid, LSST and the Nancy Grace Roman Space Telescope, for discovering new stellar streams around dwarf galaxies.

\begin{acknowledgements} 
We thank the anonymous referee, F. Prada and J. Read for insightful comments which helped improve the manuscript. JS acknowledges financial support from project PID2022-138896NB-C53 and the Severo Ochoa grant CEX2021-001131-S funded by MCIN/AEI/ 10.13039/501100011033. DMD thanks financial support for a visiting researcher stay at the Astronomy and Astrophysics Department of the University of Valencia within the framework of the «Talent Attraction» programme implemented by the Office of the Vice-Principal for Research (INV25-01-15). TS gratefully acknowledge the support of the NSF-Simons AI-Institute for the Sky (SkAI) via grants NSF AST-2421845 and Simons Foundation MPS-AI-00010513. TS was supported by NSF through grant AST-2510183 and by NASA through grants 22-ROMAN22-0055 and 22-ROMAN22-0013.

\end{acknowledgements}

\bibliographystyle{aa}
\bibliography{Bibliography}

\begin{appendix}

\section{The diversity of satellite accretion in the dwarf galaxy regime}
\label{appendix:B}

\begin{table*}[ht]
\centering
\caption{Galaxy properties, ordered by appearance in Figure \ref{fig:mosaic}. Distances and masses taken from the 50 MGC catalogue.}
\begin{tabular}{lccccccc}
\hline
\textbf{Name} & \textbf{RA} & \textbf{DEC} & \textbf{Distance (Mpc)} & \textbf{log(M$_\star$/M$_\odot$)} & \textbf{Footprint} & \textbf{Classification} \\
\hline
ESO 508-059 & 200.48 & -25.49 & $27.373 \pm 5.844$ & $8.757 \pm 0.188$ & DECaLS & Stream \\
ESO 358-054 & 55.76 & -36.27 & $9.268 \pm 5.459$ & $8.049 \pm 0.587$ & DES & Asymmetric stellar halo \\
PGC 46382 & 199.67 & -8.45 & $14.544 \pm 7.640$ & $8.569 \pm 0.507$ & DECaLS & Asymmetric stellar halo \\
UGC 300 & 7.52 & 3.51 & $16.172 \pm 4.705$ & $7.907 \pm 0.260$ & DES & Asymmetric stellar halo \\
ESO 146-002A & 329.48 & -60.31 & $19.230 \pm 6.245$ & $8.036 \pm 0.293$ & DES & Asymmetric stellar halo \\
PGC 15625 & 69.10 & -9.51 & $31.388 \pm 4.909$ & $8.602 \pm 0.137$ & DECaLS & Asymmetric stellar halo \\
PGC 15214 & 67.19 & -12.51 & $23.478 \pm 6.128$ & $8.464 \pm 0.232$ & DECaLS & Asymmetric stellar halo \\
PGC 010334 & 40.93 & -6.65 & $15.355 \pm 6.063$ & $8.834 \pm 0.363$ & DES & Asymmetric stellar halo \\
UGC 6018 & 163.52 & 20.64 & $21.905 \pm 8.676$ & $8.619 \pm 0.364$ & DECaLS & Asymmetric stellar halo \\
PGC 40604 & 186.47 & 5.81 & $16.500 \pm 1.100$ & $8.060 \pm 0.136$ & DECaLS & Shell \\
PGC 39265 & 184.00 & 4.65 & $34.600 \pm 5.333$ & $8.740 \pm 0.136$ & DECaLS & Shell \\
PGC 9774 & 38.49 & -6.36 & $15.395 \pm 5.894$ & $7.818 \pm 0.350$ & DES & Shell \\
ESO 236-041 & 326.45 & -48.80 & $17.967 \pm 5.270$ & $8.274 \pm 0.262$ & DES & Shell \\
ESO 149-003 & 358.01 & -52.58 & $7.010 \pm 0.413$ & $7.095 \pm 0.136$ & DES & Shell \\
ESO 482-013 & 54.22 & -24.91 & $28.218 \pm 7.607$ & $8.702 \pm 0.240$ & DES & Shell \\
NGC 3952 & 178.42 & -4.00 & $23.553 \pm 7.871$ & $9.274 \pm 0.302$ & DECaLS & Shell \\
PGC 755691 & 52.23 & -27.39 & $17.895 \pm 8.602$ & $7.834 \pm 0.455$ & DES & Shell \\
PGC 131167 & 47.71 & -41.80 & $22.360 \pm 9.544$ & $8.282 \pm 0.396$ & DES & Shell \\
NGC 4765 & 193.31 & 4.46 & $5.650 \pm 1.126$ & $7.883 \pm 0.175$ & DECaLS & Shell \\
NGC 4928 & 195.75 & -8.08 & $24.328 \pm 7.571$ & $9.436 \pm 0.280$ & DECaLS & Shell \\
\hline
\end{tabular}
\end{table*}

In this Appendix, we provide a more detailed description of the features identified in Figure \ref{fig:mosaic}. We mention works where tidal debris has been mentioned- all but three (PGC 40604, ESO 149-003, NGC 4765) of these stellar stream features are new discoveries. 

In the top panel, we showcase a stream (ESO 508-059) followed by asymmetric stellar halos (ESO 358-054 to ESO 146-002A). ESO 508-059 represents the most confident stream discovery during our pilot study. ESO 358-054 has an asymmetric stellar halo could be the phase-mixed remnants of a stellar stream- there is a stream-like overdensity in the east, with a "gap" between the overdensity and the galaxy. PGC 46382 has two stream-like overdensities (southeast, east), and the galaxy appears to have two stellar overdensities, with no obvious stellar core. UGC 300 has a very faint stream-like overdensity (west). ESO 146-002A's stellar halo clearly traces the past infall of a satellite. While \citealt{Makarov2014} reported a faint, very diffuse curved tail towards the NW, no deep imaging nor mention of the feature's tidal nature was provided. 

In the second panel, we show more asymmetric stellar halos. PGC 15625 showcases a satellite dwarf galaxy clearly disrupted into a stellar stream. PGC 15214 has a faint, diffuse stellar stream like overdensity (south) which is denser towards the west. In some ways, PGC 15214 could be considered a Magellanic analogue due to its LMC-like morphology, and an infalling satellite (north) which appears to be 1/10th of its size. The foreground star makes it unclear whether or not the satellite is showing a tidal tail in the direction of PGC 15214. PGC 10334 has a similar, phase-mixed southern structure, with a western overdensity, which could have been the impact site of the accretion event. The southern part of the galaxy's body is clearly pulled towards the impact site, and thus may contain the progenitor satellite. UGC 6018 shows a northwestern overdensity in its stellar halo. PGC 40604 is a blue compact dwarf (BCD) galaxy. It showcases both an asymmetric stellar halo, which clearly traces the path of the infalling satellite, and shell-like features at both ends of the galaxy. After \cite{Paudel2018} identified it as a merging galaxy, \cite{Zhang2020a, Zhang2020b} confirmed it as a coalesced gas-rich dwarf-dwarf merger (1:2 to 1:5). Using $N$-body hydrodynamical simulations to, they found that the progenitors had their first passage on a near radial, non-coplanar orbit over 1 Gyr ago and the impact triggered a central starburst. 

In the third panel we showcase shell-like features found in the SSLS. PGC 39265 has shell-like overdensities in the south-east and north-west, and a single overdensity in the north-east. PGC 9774 has symmetric double shells south and north of the galaxy. In the north-east there is a single overdensity, which could either be part of the stream or a satellite. ESO 236-041 has symmetric shells east and west of the galaxy. The main body appears warped in the direction of the shells. ESO 149-003 has already been confirmed to have shells in previous r-band images (\citealt{Ryan-Weber2004, Koribalski2018}). Its anomalous H I gas kinematics can be explained by a past accretion event or, given no remnant of the interaction partner has yet been identified, a pristine gas accretion (\citealt{Josza2021}). ESO 482-013 showcases a prominent shell in the east, with a significantly fainter second shell in the west. 

In the last panel we show more examples of shell-like features found. NGC 3952 has a shell-like overdensity in the east.  PGC 755691 has a bright, compact blue core, with (more diffuse) symmetric shells and a denser shell in the south. PGC 131167 has a distinct shell in the north, and a bright, compact core. NGC 4765 has a more saturated blue core, with a shell east of the galaxy also identified in \cite{Paudel2018}. Facing the shell is an overdensity, and it is not clear if it is at the same distance as the galaxy. Finally, NGC 4928 is a spiral dwarf galaxy with a bright blue core and symmetric shells north and south of the galaxy.

\section{Challenges in identifying stellar streams around dwarf galaxies}
\label{appendix:A}

In this Appendix we describe the observational difficulties associated with distinguishing stellar streams around dwarf galaxies.

The overarching aim of the SSLS is to study satellite accretion around nearby galaxies that have not been strongly perturbed by major merger events in the recent past. For massive galaxies, we define tidal streams as dwarf remnants, i.e., minuscule mergers ($\leq$1:3 total mass ratio). In this context, features \textit{not} related to satellite accretion in massive galaxies can arise from (i) galaxies undergoing tidal disruption within massive galaxy clusters or ongoing 1:1 major mergers, or (ii) tidal debris and tails associated with past major merger remnants (e.g., \citealt{Toomre1972}). By applying an isolation criteria we remove most of (i). For (ii), by measuring the width of the tidal features we can identify potential major merger debris, given they have been shown to scale with the progenitor mass and dynamical age (e.g., \citealt{Johnston2001, Erkal2016, MD2023}). Another possible source of confusion comes from so-called galactic feathers- i.e., perturbed stellar disc material arising from interactions with low-mass satellites giving rise to tidal tails (\citealt{Laporte2019}). While it remains to be seen if colour information alone is sufficient to reliable distinguish feathers from streams, simulations of galactic feathers in minor mergers provide a useful tool in identifying potential galactic feather contaminants (see, e.g. figure 9 of \cite{MD2023} for example simulations and their figure 10 for galactic feather candidates). Finally, clouds of dust and gas in our MW (Galactic cirri) can mimic LSB features. To minimise its presence it is recommended to study galaxies outside the zone of avoidance (Galactic latitudes $|b| < 20\degree$). Visual inspection is usually sufficient in identifying cirri, and its colours help distinguish cirri from streams (\citealt{roman2020, MD2023}).

Extrapolating the above to the dwarf galaxy regime is challenging, given dwarf galaxies are less massive and fainter. With increasing distance it becomes more challenging to reach sufficient resolution at which dwarf satellite disruption can be confidently identified. First, it becomes harder to distinguish major and minor mergers in the dwarf-galaxy regime. Depending on the total (major) merger mass ratio and angle of approach, tidal debris pulled from the host dwarf galaxy itself risks misidentification as tidal debris from satellite disruption. Tidal distortion of the host dwarf galaxy and in-situ star formation can lead to complex morphologies giving rise to stream-like components, such as extensive spiral arms. For example, PGC 15625 (Figure \ref{fig:mosaic}) appears highly disrupted as it is accreting its satellite galaxy. While we do not know where the hosts' debris material ends, we can visually see that the progenitor satellite is also highly disrupted. Another example is PGC 15214- while we are confident PGC 15214 shows satellite accretion (thanks to the diffuse track-like feature in its asymmetric stellar halo), the spiral arms north/south of the central body appear stream-like. Second, as discussed earlier, due to the SB cut off we can only detect the brightest pieces of streams. Without deeper imaging we do not know the full extent of the tidal feature- our most confident stream detection, ESO 508-059, only shows the brightest piece of its stream. Finally, the line-of-sight perspective further complicates our morphological interpretation of the streams. This is especially problematic when the stellar stream is projected onto the main body of the host and looks attached to it.  

For both of these cases, one solution would be to get kinematical data for these features- unfortunately this is unavailable for unresolved systems. Sometimes colour information can help, given spiral arms tend to be bluer due to active star formation. However, we can also expect tidal streams with bluer colour if the progenitor was a gas-rich dIrr galaxy.

Finally, tidal interactions with nearby galaxies can also produce tidal features not related to accretion. As described in Section \ref{sec:meth}, we have removed dwarf galaxies that are near a massive host or in an interacting pair. However, we have found some cases of dwarf galaxies which show tidal features (with no accretion track) and appear to have nearby dwarf galaxies (that are either satellites or foreground galaxies) close to the feature. Unfortunately, given distance information is not available for all dwarf galaxies, it is difficult to know in these cases whether these dwarfs caused the tidal feature.

\end{appendix}

\end{document}